\documentclass[conference]{IEEEtran}
\usepackage{amsmath}
\usepackage{amssymb}
\usepackage[dvips]{graphicx}
\usepackage{epsfig}

\begin{document}


\title{Energy Efficiency and Goodput Analysis in Two-Way Wireless Relay Networks}



%
\author{\authorblockN{Qing Chen and Mustafa Cenk Gursoy}
\authorblockA{Department of Electrical Engineering\\
University of Nebraska-Lincoln, Lincoln, NE 68588\\ Email:
chenqing@huskers.unl.edu, gursoy@engr.unl.edu}}


\maketitle

\begin{abstract}\footnote{This work was supported by the National Science Foundation under Grants CCF -- 0546384 (CAREER) and CNS--0834753.}
In this paper, we study two-way relay networks (TWRNs) in which  two
source nodes exchange their information via a relay node indirectly
in Rayleigh fading channels. Both Amplify-and-Forward (AF) and
Decode-and-Forward (DF) techniques have been analyzed in
the TWRN employing a Markov chain model through which the network operation is described and investigated in depth. Automatic Repeat-reQuest (ARQ)
retransmission has been applied to guarantee the successful packet
delivery. The bit energy consumption and goodput expressions have
been derived as functions of transmission rate in a given AF or DF
TWRN. Numerical results are used to identify the optimal transmission rates
where the bit energy consumption is minimized or the goodput is
maximized. The network performances are compared in
terms of energy and transmission efficiency in AF and DF modes.
\end{abstract}
\vspace{-.1cm}
\section{Introduction}
Recently, there has been much interest on two-way relay
networks (TWRNs) in which two source nodes $T_{1}$ and $T_{2}$ without a direct
link communicate with each other via  a relay node.
The architecture of TWRNs makes it possible to better exploit the
channel multiplexing of uplink and downlink wireless medium
\cite{PPHY}. The source nodes initially send their data to the relay node. The received data is combined employing a certain method
according to the Amplify-and-Forward (AF) or the Decode-and-Forward (DF)
mode and gets broadcasted from the relay back to both source nodes. With the
application of network coding and channel estimation techniques
\cite{FRY}, $T_{1}$ and $T_{2}$ can perform self-interference cancelation and
remove their own transmitted codewords from the received signal. Four time slots needed in a
traditional one-way transmission for the forward and backward
channels to accomplish one-round information exchange between
$T_{1}$ and $T_{2}$ via the relay node can be reduced to two in TWRNs by
comparison.

In a realistic multi-user wireless network, e.g. IS-856 system
\cite{DP} which has more relaxed delay requirements, the transmission
power is fixed while the rate can be adapted according to the
channel conditions. Moreover, Automatic Repeat reQuest (ARQ)
techniques have been applied to improve the transmission reliability
above the physical layer \cite{WDB}. Prior works \cite{PN}, \cite{RPC}
also show there is a compromise between transmission rate $R$ and
ARQ such that the network average successful throughput, i.e., the
goodput, can be maximized at an optimal rate $R^*$. In addition to the goodput analysis, we in this paper are interested in energy-efficient operation. In such cases,  the energy
consumption due to retransmission should also be taken into account to
evaluate the energy efficiency with respect to $R$. Hence, we investigate the joint optimization of $R$ (at the physical layer) and
the number of ARQ retransmissions (at the data-link layer) by
adopting a cross-layer framework in TWRNs.

The remainder of this paper is organized as follows. Section II
introduces the TWRN model, channel assumptions as well as its
general working mechanism. In Sections III and IV, we investigate the
Markov chain model under both AF and DF modes to derive the
analytical expressions for the bit energy consumption and
goodput. In Section V, the numerical results are shown
to compare the system performance in AF and DF modes. Section
VI provides the conclusion.


\vspace{-.2cm}
\section{System Formulation and Channel Assumptions}
\begin{figure}
\begin{center}
\includegraphics[width = 0.3\textwidth]{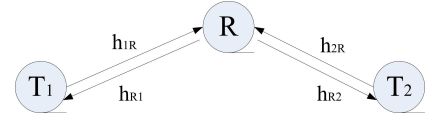}
\caption{Two-Way Relay Network} \label{fig:TWRN}
\end{center}
\end{figure}
In Figure 1, we depict a 3-node TWRN where source nodes $T_{1}$
and $T_{2}$ can only exchange information via the relay node.
Codewords $x_{1}$ and $x_{2}$ from $T_{1}$ and $T_{2}$, respectively, have equal length
and unit energy. All nodes are working in half-duplex mode and the
channels between $T_{1}$ and the relay, and $T_{2}$ and the relay are
modeled as complex Gaussian random variables with distributions
$h_{1r}\sim{\mathcal{CN}(0,\sigma_{1}^{2})}$ and
$h_{2r}\sim{\mathcal{CN}(0,\sigma_{2}^{2})}$. Without loss of
generality, we also assume channel reciprocity such that
$h_{r1}$ and $h_{r2}$ have identical distributions as
$h_{1r}$ and $h_{2r}$, respectively. Odd and even time slots have equal length, which
is the time to transmit one codeword, and are dedicated for uplink
and downlink data transmissions, respectively.  ACK and NACK control
packets are assumed to be always successfully received and the
trivial processing time is ignored. Additive Gaussian noise at the receiver terminals is
modeled as $n\sim{\mathcal{CN}(0,\sigma_{n}^{2})}$.

There are two more key assumptions: 1) channel codes support
communication at the instantaneous channel capacity levels, and
outages, which occur if transmission rate exceeds the instantaneous
channel capacity, lead to packet errors and are perfectly detected
at the receivers; 2) depending on whether packets are successfully
received or not, ACK or NACK control frames are sent and received
with no errors. Based on above network formulations, we can further
discuss the TWRN working procedure according to the current network
states under AF and DF relay schemes, and find out the inherent
impact of the transmission rate $R$ on network performances.

\section{Amplify-and-Forward TWRN}
\subsection{Network Model}
The TWRN in AF mode can be visualized as two bi-directional cascade
channels where in the odd time slots, $T_{1}$ and $T_{2}$ send
individual codewords simultaneously to the relay and the signals are
actually superimposed in the wireless medium. The relay will then
amplify the received signals proportional to the average received
power and broadcast the combined signals back to $T_{1}$ and $T_{2}$
in the even time slots.

According to Fig.\ref{fig:TWRN}, the received signals at the relay in
odd time slots is

\begin{align}
y_{r}=\sqrt{P_{1}}h_{1r}x_{1}+\sqrt{P_{2}}h_{2r}x_{2}+n,
\end{align}
where $P_{1}$, $P_{2}$ are the transmit power of $T_{1}$ and $T_{2}$
respectively. The relay will forward $y_{r}$ with a scaling factor
$\beta$ which is
\begin{align}
\beta=\frac{\sqrt{P_{r}}}{\sqrt{P_{1}\sigma_{1}^{2}+P_{2}\sigma_{2}^{2}+\sigma_{n}^{2}}}.
\end{align}
where $P_{r}$ is the relay's transmit power.

Here, we normalize the variance of the channel between $T_{1}$ and
$T_{2}$ as $\sigma^2=1$ and by using a normalized distance factor
\cite{CCN} $k=\frac{d_{T_{1},Relay}}{d_{T_{1},T_{2}}} \in(0,1)$
while assuming the variances of the other links are proportional to
$d^{-\alpha}$ where $\alpha$ is the path loss coefficient, we then have $\sigma_{1}^2=k^{-\alpha}\sigma^2$ and
$\sigma_{2}^2=(1-k)^{-\alpha}\sigma^2$. At the end of even time
slots, the received signals on $T_{1}$ and $T_{2}$ can be written as
\begin{align}
y_{1}&=\sqrt{P_{1}}\beta h_{r1}h_{1r}x_{1}+\sqrt{P_{2}}\beta
h_{r1}h_{2r}x_{2}+\beta h_{r1}n+n_{1}\nonumber\\
y_{2}&=\sqrt{P_{1}}\beta h_{r2}h_{1r}x_{1}+\sqrt{P_{2}}\beta
h_{r2}h_{2r}x_{2}+\beta h_{r2}n+n_{2},
\end{align}
where $n_{1}$, $n_{2}$, and $n$ are i.i.d Gaussian noise components. Assuming the
instantaneous channel state information is perfectly known at
$T_{1}$ and $T_{2}$, the self-interference part can be removed from $y_{1}$
and $y_{2}$ and the signals for decoding can be represented by
\begin{align}
\label{eq:AF-SIC}
 \widehat{y}_{1}&=\sqrt{P_{2}}\beta
h_{r1}h_{2r}x_{2}+\beta h_{r1}n+n_{1}\nonumber\\
\widehat{y}_{2}&=\sqrt{P_{1}}\beta h_{r2}h_{1r}x_{1}+\beta
h_{r2}n+n_{2}.
\end{align}

The cascade channel instantaneous rate from $T_{1}$ to $T_{2}$ and from
$T_{2}$ to $T_{1}$ are hence represented by
\begin{align}
R_{12}&=\log\left(1+\frac{|\beta h_{r2}h_{1r}|^2P_{1}}{(1+|\beta
h_{r2}|^2)\sigma_{n}^2}\right)\nonumber\\
R_{21}&=\log\left(1+\frac{|\beta h_{r1}h_{2r}|^2P_{2}}{(1+|\beta
h_{r1}|^2)\sigma_{n}^2}\right).
\end{align}

To describe the network mechanism more accurately, we need to
formulate the protocol of TWRN in AF mode as follows:
\begin{enumerate}
\item Each transmission round contains two consecutive time slots. In the odd slot, source nodes $T_{1}$ and $T_{2}$ both transmit
codewords to the relay with transmission rate $R$ bit/sec/Hz, and
the relay in the following even slot broadcasts $\beta y_{r}$ back to
source nodes.
\item At the end of one transmission round, $T_{1}$ and $T_{2}$ perform self-interference
cancelation (SIC) to subtract their own weighted messages, and
decode $\widehat{y}_{1}$ and $\widehat{y}_{2}$, respectively. If the
decoding fails, an outage event will be declared on that cascade
link.
\item The outage event on the cascade link $T_{1}-T_{2}$ (or $T_{2}-T_{1}$) is
defined as the probability of the event $R_{12}<R$ (or $R_{21}<R$). ACK or
NACK packets would be sent back to the relay based on successful
transmission or outage. The relay will also notify $T_{1}$ and $T_{2}$
whether a new codeword or an old codeword should be (re)transmitted
in the next odd time slot with the control packet information.
\end{enumerate}

\begin{figure}
\begin{center}
\includegraphics[width = 0.2\textwidth]{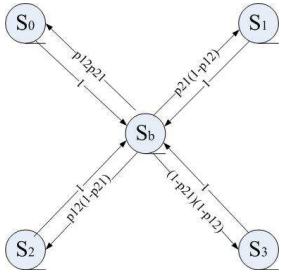}
\caption{Markov Chain of TWRN in AF} \label{fig:AF-Markov}
\end{center}
\end{figure}

The network state transition diagram of the AF TWRN can be modeled
as a Markov chain as shown in Fig. \ref{fig:AF-Markov}, where the probability
on each path denotes the probability of  the transition between two
states. $p_{12}$ and $p_{21}$ are defined as the outage
probabilities on the cascade $T_{1}-T_{2}$ and $T_{2}-T_{1}$ links and are given by
\begin{align}
p_{12}&=p(R_{12}<R)=p\left(\frac{|h_{r2}h_{1r}|^2}{\frac{1}{\beta^2}+|h_{r2}^2|}<\frac{(2^{R}-1)\sigma_{n}^2}{P_{1}}\right)\label{eq:AF-Outage0}\\
p_{21}&=p(R_{21}<R)=p\left(\frac{|h_{r1}h_{2r}|^2}{\frac{1}{\beta^2}+|h_{r1}^2|}<\frac{(2^{R}-1)\sigma_{n}^2}{P_{2}}\right)\label{eq:AF-Outage}.
\end{align}
(\ref{eq:AF-Outage0}) and (\ref{eq:AF-Outage}) can be determined using the cumulative
distribution function of the random variable $X$ \cite{AYM}
\begin{align}
F_{X}(x)&=1-\frac{1}{\mu_{2}}\int_{0}^{\infty}e^{-\frac{x(a+z)}{\mu_{1}z}-\frac{z}{\mu_{2}}}dz\intertext{where}
X&=\frac{Y_{1}Y_{2}}{a+Y_{2}},
\end{align}
and $Y_1$ and $Y_{2}$ are independent exponential distributed with
mean $\mu_{1}$ and $\mu_{2}$, and $a$ is a constant. In this context,
we know
\begin{align}
\left\{
\begin{array}{l}
\mu_{1}=E[|h_{1r}|^2],\,\hspace{0.3cm}\mu_{2}=E[|h_{r2}|^2], \hspace{0.3cm} for \hspace{0.3cm} p_{12}\\
\mu_{1}=E[|h_{2r}|^2],\,\hspace{0.3cm}\mu_{1}=E[|h_{r1}|^2], \hspace{0.3cm} for \hspace{0.3cm} p_{21}\\
\alpha=\frac{1}{\beta^2}
\end{array}
\right.
\end{align}
With the given network parameters, $p_{12}$ and $p_{21}$ can be
derived accordingly.

\subsection{Goodput Analysis}
\begin{figure}
\begin{center}
\includegraphics[width = 0.2\textwidth]{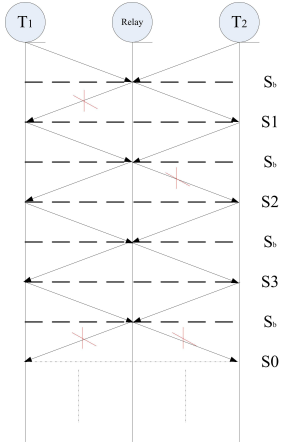}
\caption{State Transition of TWRN in AF} \label{fig:AF-State}
\end{center}
\end{figure}

In Fig. \ref{fig:AF-State}, it is explicitly seen at the beginning of
each odd time slot that both $T_1$ and $T_2$ transmit to the relay such that
at the beginning of each even time slot, the relay is always in the
ready-for-broadcasting state $S_{b}$ and will consequently transition
to $S_{0}$, $S_{1}$, $S_{2}$, or $S_{3}$ with certain probabilities.
We first determine the following four equations according to the state transitions in the
Markov chain to derive the probability of each state:
\begin{align}
\label{eq:AF-markov-equations}
\begin{cases}
p(S_{0})=p(S_{b})p_{12}p_{21}\\
p(S_{1})=p(S_{b})p_{21}(1-p_{12})\\
p(S_{2})=p(S_{b})p_{12}(1-p_{21})\\
p(S_{3})=p(S_{b})(1-p_{21})(1-p_{12})\\
p(S_{0})+p(S_{1})+p(S_{2})+p(S_{3})+p(S_{b})=1.
\end{cases}
\end{align}
After solving the set of equations in (\ref{eq:AF-markov-equations}), we obtain
$p(S_{b})=\frac{1}{2}$. We know from the inherent characteristics
of the AF TWRN that the data exchange only happens in the broadcasting
phases with successful packet delivery. Therefore, the system
goodput is defined similarly as in \cite{PPHY} by
\begin{align}
\label{eq:AF-goodput}
&\eta_{AF}\nonumber\\
&=p(S_{b})R(2(1-p_{12})(1-p_{21})+p_{12}(1-p_{21})+p_{21}(1-p_{12}))\nonumber\\
&=\frac{R(2-p_{12}-p_{21})}{2}.
\end{align}
(\ref{eq:AF-goodput}) indicates through the terms $p_{12}$ and $p_{21}$ that a higher transmission
rate $R$ will result in higher packet error rates (outage), leading
to more ARQ retransmissions which equivalently reduce the data rate.
Intuitively, a balance between transmission rate and the number of ARQ
retransmissions needs to be found such that the goodput is maximized.

\subsection{Average Bit Energy Consumption}
Energy efficiency has always been a major concern in wireless
networks. Recently, power or energy efficiency in
wireless one-way relay networks have been extensively studied. In
\cite{QM}, the average bit energy consumption $E_{b}$ is
minimized by determining the optimal number of bits per symbol, i.e., the
constellation size, in a specific modulation format. Similarly as in one-way relay channels, the
outage probabilities in TWRNs are functions of the transmission rate $R$.
For instance, there could be an increased number of outage events on the cascade channels when
codewords are transmitted at a high rate. In such a case, more
retransmissions and higher energy expenditure are needed to accomplish the
reliable packet delivery. Therefore, we are interested in a possible
realization of TWRN operation, which can provide a well-balanced performance
on both the goodput $\eta$ and the required energy. We evaluate this by
formulating the average bit energy consumption $E_{b}$ required for
successfully exchanging one information bit between $T_{1}$ and
$T_{2}$.

We then evaluate $E_{b}$ by considering long-term transmissions on
TWRN. Regardless of the previous state, whenever the relay is in state
$S_{b}$ and is broadcasting, the resulting state would be any of the other four
states previously described. Assuming there are $K$ rounds of
two-way transmission, each of which consists of a pair of
consecutive time slots and each codeword has $L$ bits. Therefore, with
$\sum_{i=0}^{3}K_{i}=K$, where $K_{i}$ is the number of transmission
round corresponding to state $S_{i}$, the average bit energy
consumption could be derived as the ratio of total bits successfully
exchanged over total energy consumption:
\begin{align}
\label{eq:AF-Eb}
E_{b}&=\lim_{K\rightarrow\infty}\frac{K(P_{1}+P_{2}+P_{r})\frac{L}{R}}{K_{3}2L+K_{1}L+K_{2}L}\nonumber\\
&=\lim_{K\rightarrow\infty}\frac{(P_{1}+P_{2}+P_{r})\frac{L}{R}}{\frac{K_{3}}{K}2L+\frac{K_{1}}{K}L+\frac{K_{2}}{K}L}\nonumber\\
&=\frac{P_{1}+P_{2}+P_{r}}{(2-p_{12}-p_{21})R}.
\end{align}

\section{Decode-and-Forward TWRN}
\subsection{Network Model}
The DF TWRN differs from the AF TWRN in that there is a crucial
intermediate decoding procedure at the the relay when it has received the
codeword from the uplink transmission. If both source nodes are allowed
to send codewords to the relay simultaneously in the uplink transmission,
the decoding at the relay has to deal with the multiple access problem in a
realistic application, with successive interference cancelation
techniques. To reduce the hardware complexity and increase the
feasibility of implementation, we hereby adopt the DF TWRN mode from
\cite{PPHY} where the relay performs sequential Decode-and-Forward. The
outage probabilities on $T_{1}-Relay$, $T_{2}-Relay$, $Relay-T_{1}$,
$Relay-T_{2}$ links are denoted as $p_{1r}$, $p_{2r}$, $p_{r1}$ and
$p_{r2}$. The protocol for sequential DF TWRN is described as
follows:
\begin{enumerate}
\item In the initial state $S_{0}$,  the relay's buffer is empty and the relay first
polls on $T_{1}$ until it receives codeword $x1$ successfully with
probability $1-p_{1r}$. Then, the state moves to $S_{1}$ which means
the relay holds $x_{1}$ in the buffer. Otherwise, the state remains as
$S_{0}$ with probability $p_{1r}$.

\item If the relay already has $x_{1}$, it starts polling $T_{2}$. The state
$S_{1}$ either changes to $S_{3}$ with probability $1-p_{2r}$ upon
successfully receiving $x_{2}$, or stays in $S_{1}$ with $p_{2r}$.

\item When the relay has both $x_{1}$ and $x_{2}$, it generates a new
codeword
$y_{n}=\sqrt{\frac{P_{r}}{2}}x_{1}+\sqrt{\frac{P_{r}}{2}}x_{2}$
according to a Gaussian codebook of $2^{2LR}$ with equal power
allocation. Then at rate $R$, it broadcasts to $T_{1}$ and $T_{2}$,
which will perform SIC to decode $x_{2}$ and $x_{1}$, respectively. Accordingly,
the state will transit to $S_{0}$, $S_{1}$, $S_{2}$ or $S_{3}$ with
corresponding probabilities $(1-p_{r1})(1-p_{r2})$,
$(1-p_{r2})p_{r1}$, $(1-p_{r1})p_{r2}$ or $p_{r1}p_{r2}$
respectively.

\item At the beginning of next transmission round, the relay will
decide to poll a new codeword from $T_{1}$ (or $T_{2}$) based on the
previous state being $S_{0}$,$S_{2}$ (or $S_{1}$) or just retransmits the old
$y_{n}$ if the previous state was $S_{3}$.
\end{enumerate}

\begin{figure}
\begin{center}
\includegraphics[width = 0.3\textwidth]{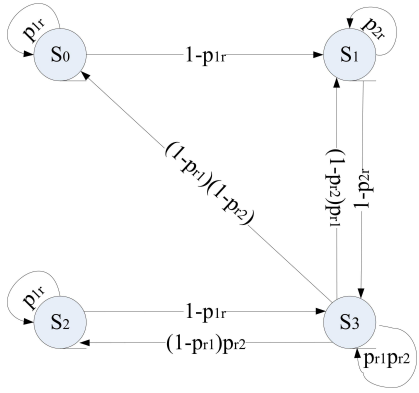}
\caption{Markov Chain of TWRN in DF} \label{fig:DF-Markov}
\end{center}
\end{figure}

The network state transition diagram of the DF TWRN can be modeled as
in Fig. \ref{fig:DF-Markov} with detailed probabilities on each
path. Since the relay receives and decodes $x_{1}$ and $x_{2}$ at
different time slots, the received signals at the relay from uplink
transmissions can be represented as
\begin{align}
y_{1r}&=\sqrt{P_{1}}h_{1r}x_{1}+n_{1}\nonumber\\
y_{2r}&=\sqrt{P_{2}}h_{2r}x_{2}+n_{2}.
\end{align}
Similar to (\ref{eq:AF-SIC}), the signals for decoding at $T_{1}$
and $T_{2}$ after SIC has been performed can be written as
\begin{align}
x_{r1}&=\sqrt{\frac{P_{r}}{2}}h_{r1}x_{2}+n_{1}\nonumber\\
x_{r2}&=\sqrt{\frac{P_{r}}{2}}h_{r2}x_{1}+n_{2}.
\end{align}

\subsection{Goodput Analysis}
Similarly as in the discussion of the goodput of AF TWRN in Section III, the data exchange
only occurs upon the successful signal receptions at $T_{1}$ and $T_{2}$
at the end of the broadcasting time slot. Hence, initially,
it is necessary to calculate the probability of being in state stays 3, i.e., $p(S_{3})$.

We start from calculating the outage probabilities on the forward
and backward channels as
\begin{align}
p_{1r}&=1-e^{\frac{-(2^R-1)\sigma_{n}^2}{\mu_1P_{1}}}\nonumber\\
p_{2r}&=1-e^{\frac{-(2^R-1)\sigma_{n}^2}{\mu_2P_{2}}}\nonumber\\
p_{r1}&=1-e^{\frac{-(2^R-1)2\sigma_{n}^2}{\mu_1P_{r}}}\nonumber\\
p_{r2}&=1-e^{\frac{-(2^R-1)2\sigma_{n}^2}{\mu_2P_{r}}}.
\end{align}
The probabilities of buffer states can be solved by noting the following relations from Fig.
\ref{fig:DF-Markov}:
\begin{align}
\label{eq:DF-markov-equations}
\begin{cases}
p(S_{0})=p(S_{0})p_{1r}+p(S_{3})(1-p_{r1})(1-p_{r2})\\
p(S_{1})=p(S_{1})p_{r2}+p(S_{3})(1-p_{r2})p_{r1}\\
p(S_{2})=p(S_{2})p_{1r}+p(S_{3})(1-p_{r1})p_{r2} \\
p(S_{3})=p(S_{3})p_{r1}p_{r2}+p(S_{1})(1-p_{2r})+p(S_{2})(1-p_{1r}).
\end{cases}
\end{align}
Solving the equations in (\ref{eq:DF-markov-equations}) with given outage
probabilities, we can obtain the following results for the buffer
states:
\begin{align}
\label{eq:AF-State-prob}
\begin{cases}
p(S_{0})&=\frac{(1-p_{2r})(1-p_{r1})(1-p_{r2})}{D}\\
p(S_{1})&=\frac{(1-p_{1r})(1-p_{r2})}{D}\\
p(S_{2})&=\frac{(1-p_{2r})(1-p_{r1})p_{r2}}{D}\\
p(S_{3})&=\frac{(1-p_{1r})(1-p_{2r})}{D},
\end{cases}
\end{align}
where the polynomial in the denominators is denoted by
\begin{align}
D&=3-2p_{1r}-2p_{2r}-p_{r1}-p_{r2}+p_{r1}p_{2r}+p_{1r}p_{r2}+p_{1r}p_{2r}.
\end{align}
Therefore, the system goodput in the DF mode can be derived as
\begin{align}
&\eta_{DF}\nonumber\\
&=p(S_{3})R \left(2(1-p_{r1})(1-p_{r2})+p_{r1}(1-p_{r2})+p_{r2}(1-p_{r1})\right)\nonumber\\
&=\frac{R(2-p_{12}-p_{21})(1-p_{1r})(1-p_{2r})}{p_{2r}p_{1r}+p_{r2}p_{1r}+p_{r1}p_{2r}+3-2p_{1r}-2p_{2r}-p_{r1}-p_{r2}}.
\end{align}

\subsection{Average Bit Energy Consumption}
$E_{b}$ in the DF TWRN is more complicated to calculate than in the
AF TWRN where each transmission round has fixed power as can be seen
in (\ref{eq:AF-Eb}). Hence, in the DF scenario, we have to separate the
energy expenditure into two parts, energy consumption in the first
stage and energy consumption in the second stage. The first stage
denotes the state transition from any of 4 previous states to state
$S_{3}$, where the relay holds two codewords $x_{1}$ and $x_{2}$ in its
buffer and is ready to broadcast. The second stage is that the relay
broadcasts its newly generated codeword and the state transits back
to any of the four states again.

Considering the relay's buffer is to be loaded with both codewords
$x_{1}$ and $x_{2}$ from any of the previous states on the first
stage, the energy consumption conditioned on the previous state
$S_{0}$, $S_{1}$, $S_{2}$, or $S_{3}$ on this particular transition
will be
\begin{align}
E_{S_{0}}&=\frac{P_{1}L}{(1-p_{1r})R}+\frac{P_{2}L}{(1-p_{2r})R}\nonumber\\
E_{S_{1}}&=\frac{P_{2}L}{(1-p_{2r})R}\nonumber\\
E_{S_{2}}&=\frac{P_{1}L}{(1-p_{1r})R}\nonumber\\
E_{S_{3}}&=0.
\end{align}
On the second stage, the energy consumption for broadcasting is
always $\frac{P_{r}L}{R}$, so the average bit energy consumption for
one information bit successfully exchanged on the DF TWRN can be
computed as
\begin{align}
&E_{b}=\frac{\sum_{i=0}^{3}(E_{S_{i}}+\frac{P_{r}L}{R})p(S_{i})}{(2(1-p_{r1})(1-p_{r2})+(1-p_{r1})p_{r2}+(1-p_{r2})p_{r1})L}\nonumber\\
&=\frac{E_{S_{0}}p(S_{0})+E_{S_{1}}p(S_{1})+E_{S_{2}}p(S_{2})+\frac{P_{r}L}{R}}{(2-p_{r1}-p_{r2})L}.
\end{align}
whenever the state probabilities and outage probabilities are known.

\section{Numerical Results and Comparisons}
In this section, we present the numerical results to evaluate the
system performance of TWRN in both AF and DF modes. The network
configurations are assumed to be as follows:  Relay is located in
the middle between $T_{1}$ and $T_{2}$ which means $k=0.5$. The
power spectrum density of the Gaussian white noise is
$\sigma_{n}^2=10^{-10}$ and the channel bandwidth is set to
$B=10^6$ Hz. Path loss coefficient is $\alpha=3.12$ \cite{VE}. We also
assume the same transmit power for both source nodes and the relay,
which is $P_{1}=P_{2}=P_{r}=P$ and define the SNR by
$\gamma=\frac{P}{\sigma_{n}^2}$.

\begin{figure}
\begin{center}
\includegraphics[width = 0.5\textwidth]{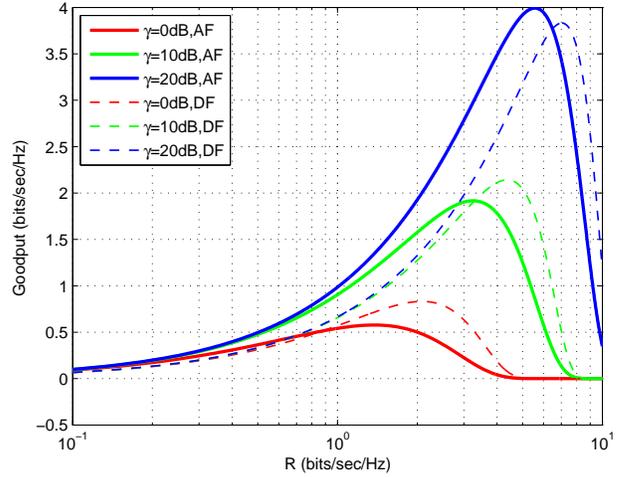}
\caption{Goodput Vs. Transmission Rate in AF and DF TWRN}
\label{fig:Eff-R}
\end{center}
\end{figure}

Firstly, we are particularly interested in how the goodput varies as a function of the transmission rate $R$ at specific SNR
values. In Fig. \ref{fig:Eff-R}, $\eta_{AF}$ and $\eta_{DF}$ are
plotted as functions of $R$, with solid and dashed lines
corresponding to AF and DF modes, respectively. On each curve with a
given specific $\gamma$ value, it's immediately seen that the
goodput first increases within low $R$ range and then begins to
drop once the rate is increased beyond the optimal $R^*$ which maximizes the goodput $\eta$.
Additionally, at low values of the rate $R$, the AF TWRN has higher goodput
$\eta_{AF}$, while beyond a certain rate $R$, DF starts to outperform,
regardless of the SNR $\gamma$ value.

\begin{figure}
\begin{center}
\includegraphics[width = 0.5\textwidth]{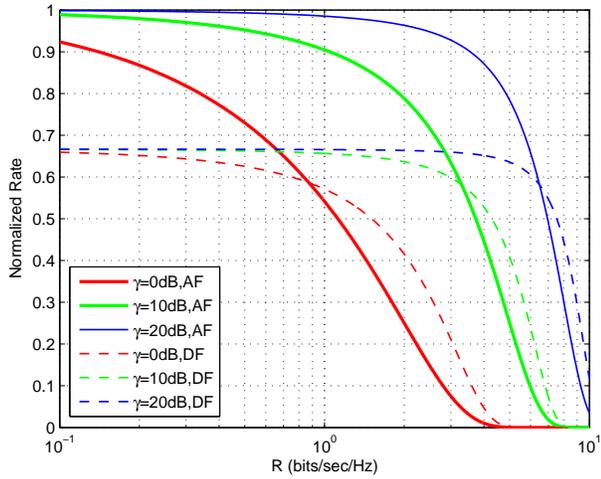}
\caption{Normalized Transmission Rate Vs. Transmission Rate}
\label{fig:Normal-R}
\end{center}
\end{figure}

To better illustrate the goodput performance in AF and DF modes, we
look into transmission efficiency  by defining a normalized rate
$\frac{\eta}{R}$ and plotting it in Fig. \ref{fig:Normal-R}. The normalized
rate is always decreasing when $R$ increases at all SNR's in both AF
and DF. In other words, increasing outage probabilities due to
increasing $R$ has eventually resulted in more ARQ retransmissions. Specifically in the
high SNR scenario, the normalized rate levels off between 0.6 and
0.7 in AF and 0.9 and 1 in DF, which means the transmission
efficiency doesn't change too much within this rate range. In
addition, the AF mode seems to have higher normalized rate in low
rate range.

\begin{figure}
\begin{center}
\includegraphics[width = 0.5\textwidth]{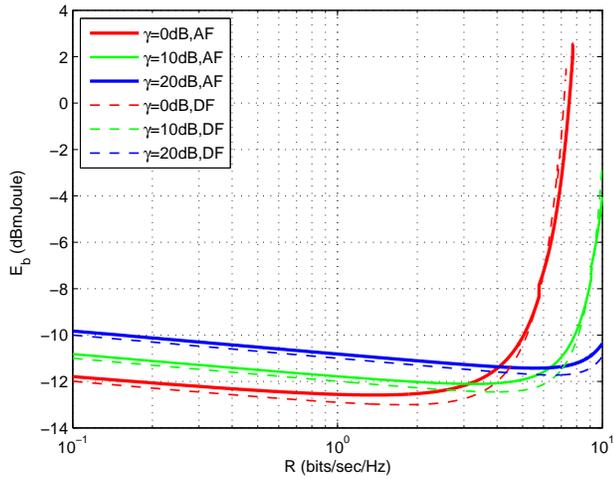}
\caption{Average Bit Energy Vs. Transmission Rate}
\label{fig:Eb-R}
\end{center}
\end{figure}

\begin{figure}
\begin{center}
\includegraphics[width = 0.5\textwidth]{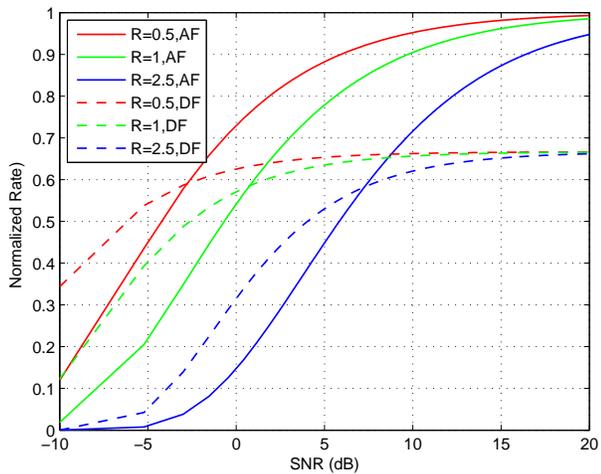}
\caption{Normalized Transmission Rate Vs. SNR}
\label{fig:Normal-SNR}
\end{center}
\end{figure}

In Fig. \ref{fig:Eb-R}, we analyze the energy efficiency. We notice that the difference of the average
bit energy consumptions between two modes is insignificant up until
$R=1$, but DF stills has a better energy efficiency with a lower
$E_{b}$ regardless of $\gamma$. However when compared with the
corresponding points ($R<1$) in Fig. \ref{fig:Normal-R}, it is shown that
even though DF can achieve a slightly lower $E_{b}$ than AF, it also
suffers a lower transmission efficiency in the metric of lower
normalized rate.  Above $R=1$ in low SNR scenario of $\gamma=0$, AF
predominates with both higher normalized rate and lower $E_b$ until
$R$ approaches about $6$ bits/sec/Hz. Similar results can be observed
on the high SNR scenarios also. In Fig. \ref{fig:Normal-SNR}, we study
the impact of SNR on the normalized rate in both AF and DF.
Basically the normalized rate is increasing as SNR increases at all
transmission rates. At low SNR e.g. $R=-5 \hspace{0.1cm}dB$, DF
performs better than AF. As SNR approaches $R=20 \hspace{0.1cm}dB$,
the normalized rate gets close to 1 (or 0.7) in AF (or DF) mode.
Consequently, one way to improve on the transmission efficiency is
to increase SNR in the TWRN. Considering overall impacts of rate $R$
and SNR, we can always find an scheme for the TWRN to achieve
optimality in respect to the goodput $\eta$, the average bit energy
consumption $E_{b}$ or the transmission efficiency.

\section{Conclusion}
In this paper, we have studied the two-way relay networks working in
Amplify-and-Forward and Decode-and-Forward modes. In each mode, we
set up a Markov chain model to analyze the state transition in
details. ARQ transmission is employed to guarantee the successful
packet delivery at the end and mathematical expressions for the
goodput and bit energy consumption have been derived. Several
interesting results are observed from simulation results: 1) the
transmission rate $R$ can be optimized to achieve a maximal goodput
in both AF and DF modes; 2) generally the transmission efficiency is
higher in AF within a certain $R$ range, while the DF can achieve a
slightly higher energy efficiency instead; 3) increasing SNR will
always increase the normalized rate regardless of $R$. Hence, it's
possible the network performance be optimized in a balanced manner
to maintain a relatively high goodput as well as a low $E_{b}$.

\end{document}